\pdfoutput=1
\documentclass[aps,twocolumn,pra,tightenlines,floatfix,showpacs,superscriptaddress]{revtex4-1}
\usepackage{graphicx}
\usepackage{epstopdf}
\usepackage[english]{babel}
\usepackage{amsmath}
\usepackage{amssymb}
\usepackage{mathtools}
\usepackage{times}
\usepackage{appendix}
\usepackage{bm}
\usepackage{float}
\usepackage{color}
\usepackage{longtable}
\usepackage{url}
\usepackage[usenames,dvipsnames]{xcolor}
\usepackage[colorlinks=true,linkcolor=Blue,urlcolor=Blue,citecolor=Blue]{hyperref}

\graphicspath{{./Figures}}

\begin{document}

\title{Multiqubit matter-wave interferometry under decoherence and the Heisenberg scaling recovery}

\author{Yanming Che}
\affiliation{Zhejiang Province Key Laboratory of Quantum Technology and Device, Department of Physics,
Zhejiang University, Hangzhou, Zhejiang 310027, China}

\author{Jing Liu}
\affiliation{MOE Key Laboratory of Fundamental Physical Quantities Measurement
\& Hubei Key Laboratory of Gravitation and Quantum Physics, PGMF and School of Physics,
Huazhong University of Science and Technology, Wuhan 430074, China}

\author{Xiao-Ming Lu}
\email{luxiaoming@gmail.com}
\affiliation{Department of Physics, Hangzhou Dianzi University,
Hangzhou, Zhejiang 310018, China}

\author{Xiaoguang Wang}
\email{xgwang1208@zju.edu.cn}
\affiliation{Zhejiang Province Key Laboratory of Quantum Technology and Device, Department of Physics,
Zhejiang University, Hangzhou, Zhejiang 310027, China}

\date{\today}

\begin{abstract}
Most matter-wave interferometry (MWI) schemes for quantum sensing have
so far been evaluated in ideal situations without noises. In this work, we provide
assessments of generic multiqubit MWI schemes under Markovian dephasing
noises. We find that for certain classes of the MWI schemes with scale factors
that are nonlinearly dependent on the interrogation time, the optimal precision
of maximally entangled probes \emph{decreases} with increasing the particle
number $N$, for both independent and collective dephasing situations. This
result challenges the conventional wisdom found in dephasing Ramsey-type
interferometers. We initiate the analyses by investigating the optimal precision
of multiqubit Sagnac atom interferometry for rotation sensing. And we show that
due to the competition between the unconventional interrogation-time quadratic
phase accumulation and the exponential dephasing processes, the
Greenberger-Horne-Zeilinger (GHZ) state, which is the optimal input state in
noiseless scenarios, leads to vanishing quantum Fisher information in the large-$N$
regime. Then our assessments are further extended to generic MWI schemes for
quantum sensing with entangled states and under decoherence. Finally, a quantum
error-correction logical GHZ state is tentatively analyzed, which could have the
potential to recover the Heisenberg scaling and improve the sensitivity.
\end{abstract}

\maketitle

\section{Introduction}

\vspace*{-1.5ex}
Matter-wave interferometry (MWI) is sensitive to
inertial effects and has been widely used in quantum sensing of physical
quantities, including gravitational force, acceleration, and rotation of
reference frames~\cite{DegenRMP}. With quantum entanglement as resources,
quantum sensing is expected to achieve higher precision and sensitivity,
e.g., the Heisenberg limit~\cite{GiovannettiScience2004,GiovannettiPRL2006}.
Sagnac atom-interferometry gyroscopes (SAIGs) are quantum sensors for rotation
frequency based on the Sagnac interferometry~\cite{BarrettCRPhys2014} of matter
waves, where atoms are coherently split and controlled with wave guides (e.g.,
see Ref.~\cite{KasevichPRL1991}) to enclose a finite area in space and encode
the rotation frequency into the Sagnac phase, which can be finally read out
from the interference fringes~\cite{KasevichPRL1991,GustavsonPRL1997,
BarrettCRPhys2014,KandesarXiv2013,TrubkoPRL2015,HelmPRL2015,HainePRL2016,HelmPRL2018}.

Most of schemes for SAIG utilize both wave nature and spin degrees of
freedom (hyperfine states) of atoms. For example, a scheme with uncorrelated
and trap-guided atomic clocks was proposed in Ref.~\cite{StevensonPRL2015}, and
was later generalized to the one with multiparticle
Greenberger-Horne-Zeilinger (GHZ) state to beat the standard quantum limit (SQL)
in Ref.~\cite{LuoPRA2017}. So far, these proposed schemes were considered in
ideal situations where the sensing protocols consisted of perfect unitary
quantum channels. However, in realistic experiments, inevitable noises may
cause errors which prevent from the expected precision.

In standard Ramsey interferometers for atomic clocks, where the transition
frequency $\omega$ between two energy levels of atoms is measured, the phase
accumulation is linear in the interrogation time while the dephasing caused by
noises is exponential. And the use of entangled states has been proved to only
give a constant improvement for the ultimate precision, but still follows the
SQL~\cite{HuelgaPRL1997,EscherNPhys2011,DemkowiczNC2012}. Several strategies
and techniques have been proposed and used to protect the precision of atomic
clocks from noises~\cite{TanPRA2013,LiuPRA2017,*LiuPRA2017MultiPara,KesslerPRL2014,DurPRL2014,UndenPRL2016}.
Whereas, the evaluation and optimization of generic multiqubit MWI schemes for
quantum sensing under decoherence still remain challenging.

In this paper, we present an assessment of generic MWI schemes
with maximally entangled states and under dephasing noises.
Start with the SAIG as a prototype, we analyze the competition between the
unconventional phase accumulation and the exponential dephasing processes.
And we find that for certain classes of the MWI schemes with scale factors
that are nonlinearly dependent on the interrogation time, the optimal precision
of maximally entangled probes \emph{decreases} with increasing the particle
number $N$, for both independent and collective dephasing situations. These
classes include most of the current mainstream MWI schemes with atomic clock
states and certain time-dependently controlled Hamiltonian systems. Our findings
challenge the conventional wisdom found in dephasing Ramsey-type
interferometers~\cite{HuelgaPRL1997,EscherNPhys2011,DemkowiczNC2012}.

This paper is organized as follows. In Sec.~\ref{sec:ModelandPhase} we model
the matter-wave Sagnac interferometry with maximally entangled
states and derive the multiparticle Sagnac phase. In Sec.~\ref{sec:SensitivityDecoherence}
we evaluate the optimal sensitivity of Sagnac-type interferometers under local
decoherence noises via the quantum Fisher information. In Sec.~\ref{sec:GenericMWI}, we
provide assessments of generic multiqubit MWI schemes with GHZ inputs and under
independent Markovian dephasing noises. In Sec.~\ref{sec:CollectiveDephasing}
the QFI of generic multiqubit MWI schemes with GHZ inputs under collective
dephasing is provided. In Sec.~\ref{sec:QEC} the potential of recovering the
Heisenberg scaling with quantum error-correction codes is presented. Finally,
in Sec.~\ref{sec:Conclusion}, we conclude our work and give further discussions
on the minor enhancement by entanglement in the precision of Sagnac-type
interferometers.

\vspace*{-1.5ex}
\section{Matter-wave Sagnac interferometry with entangled states}
\vspace*{-1.5ex}
\label{sec:ModelandPhase}
In order to sense the rotation frequency of a reference frame ${\cal{R}}$ rotating in an
angular velocity $\bf \Omega$ with respect to an inertial frame ${\cal{K}}$,
$N$ two-state cold atoms~\cite{NumberNote} can be initially prepared at the
GHZ state (e.g., via the nonlinear~\cite{WinelandPRA1996,LeibfriedScience2004,MolmerPRL1999,LeePRL2006,GietkaPRA2015,PezzePRL2009}
or Rydberg blockade~\cite{SaffmanPRL2009} interactions with suitable coupling
parameters), which is the optimal multiparticle input state for unitary quantum
channels~\cite{PangPRA2014}, i.e., $|\psi_0\rangle=(|{\bf 0}\rangle+|{\bf 1}\rangle)/\sqrt{2}$,
where $|{\bf 0}\rangle=\left|0\right\rangle ^{\otimes N}$ and
$|{\bf 1}\rangle=\left|1\right\rangle ^{\otimes N}$, with
$\left|0\right\rangle = \left|\uparrow\right\rangle \left(\left|1\right\rangle = \left|\downarrow\right\rangle\right)$
being the single-atom (pseudo)spin state and eigenstate of Pauli matrix
$\sigma_z$ with eigenvalue $+1$ ($-1$). Subsequently, the $|{\bf 0}\rangle$ and
$|{\bf 1}\rangle$ components are coherently split by a beam splitter which
establishes a state-path entanglement, and then are guided to transport along
two distinct paths in real space~\cite{BordePLA1989}, within an interrogation
time $\tau$, and are finally recombined at $t=\tau$. The state-path entanglement
associates the phase shift between two interferometer paths (Sagnac phase) with
the relative phase of two atomic states (qubit phase), which can be
read out from the atomic spectroscopy, e.g., parity measurement~\cite{WinelandPRA1996,GerryPRA2010,LuoPRA2017},
after applying a $\pi/2$ pulse.

\begin{figure}
\centerline{\includegraphics[height=1.1in,width=3.3in,clip]{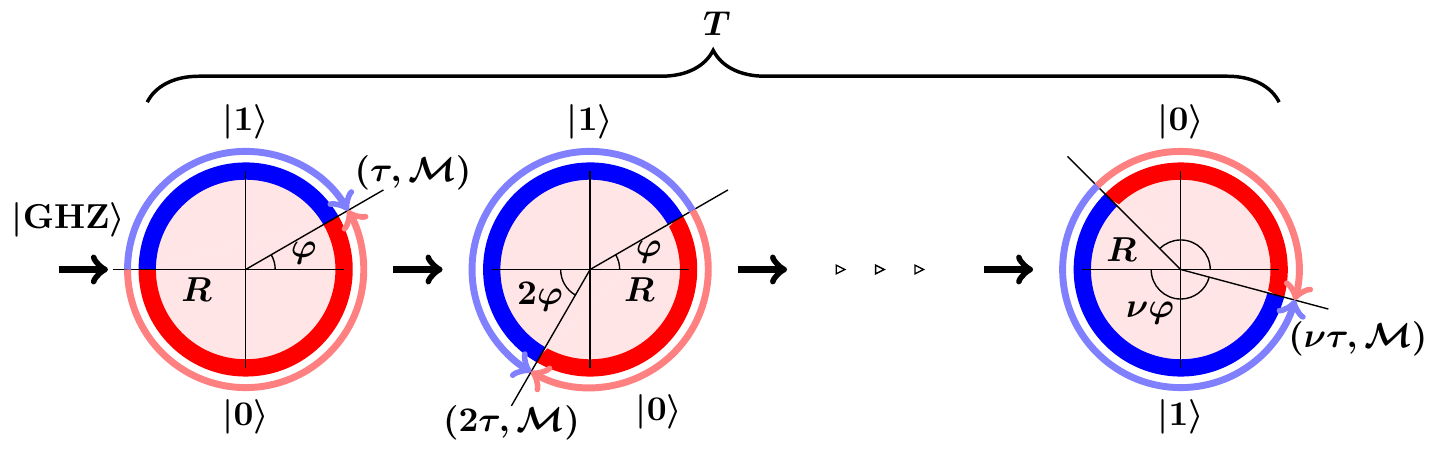}}
\caption{(Color online) Schematic matter-wave Sagnac interferometer for rotation sensing with
entangled GHZ input of BEC atoms and circular waveguides of radius $R$ in the $x-y$ plane, observed
from the inertial frame ${\cal{K}}$. The $|{\bf 0}\rangle$ (red) and $|{\bf 1}\rangle$ (blue)
components are coherently split at $t=0$ and then are counter-transported along the two
interferometer paths. Finally they are recombined at time $t=\tau$, where a $\pi/2$ pulse and
measurement ${\cal M}$ are performed to read out the encoded rotation frequency. The above process
is repeated for $\nu=T/\tau$ times to achieve a high precision. The tuple at the end of each period
denotes (time, measurement). The reference frame rotates with an angular velocity
${\bf{\Omega}}=\Omega{\bf{z}}$ and $\varphi=\Omega \tau$.}
\label{fig:SI}
\end{figure}

\emph{Model.---}We assume that the $N$ two-state bosonic atoms are in the
Bose-Einstein condensed (BEC) state, which is described by the mean-field wave
function (order parameter) $\Psi_{\xi}\left({\bf r}, t\right)$ for the two split
components $\left|\xi\right\rangle = \left|0\right\rangle$ and $\left|\xi\right\rangle = \left|1\right\rangle$,
respectively. And the wave guide is provided by a ring trap of toroidal
geometry~\cite{HalkyardPRA2010,KandesarXiv2013,HelmPRL2015,HainePRL2016},
with a trapping potential in cylindrical coordinates $\left\{r, \theta, z\right\}$
of the form $V_{\mathrm{trap}}\left({\bf r}, t\right)=\frac{1}{2}m
\left[\omega_r^2\left(r-R\right)^2+\omega_{\theta}^2 R^2 \theta^2 \Theta(-t)
+\omega_z^2 z^2\right]$~\cite{KandesarXiv2013,HainePRL2016},
where $m$ is the particle mass and ($\omega_r, \omega_{\theta}, \omega_z$) are
the respective (radial, angular, axial) trapping frequencies, and $\Theta(-t)$
and $R$ are the Heaviside step function and the radius of the circular
interferometer, respectively. See Fig.~\ref{fig:SI} for a schematic
illustration, where we assume ${\bf{\Omega}}=\Omega{\bf{z}}$. When the radial
and axial trapping confinements are sufficiently tight, the dynamics along
these directions is freezed and then the time evolution ($t \ge 0$) of the
order parameter in the rotating frame ${\cal{R}}$ is given by the one-dimensional
Gross-Pitaevskii (GP) equation~\cite{DalfovoRMP1999}
\begin{equation}
\label{eq:GPE}
i\hbar \frac{\partial}{\partial t}\Psi_{\xi}\left(\theta, t\right)
 = H_{\xi}\Psi_{\xi}\left(\theta, t\right),
\end{equation}
with the mean-field Hamiltonian
\begin{equation}
\label{eq:GPEHamiltonian}
H_{\xi} = \frac{\hat{L}_z^2}{2mR^2} +
{\cal U} \left| \Psi_{\xi}\left(\theta, t\right)\right|^2 - \Omega \hat{L}_z,
\end{equation}
where $\hat{L}_z=-i\hbar \frac{\partial}{\partial \theta}$ is the axial angular
momentum operator and ${\cal U}$ is the contact interaction strength.

For the ${\cal U}=0$ case, the time evolution operator for the $i$th particle
reads
\begin{equation}
\label{eq:Uoperator}
\hat{U}_i\left(t\right)=\mathrm{exp}\left(\Omega t \frac{\partial}
{\partial \theta_i}\right)\mathrm{exp}\left[\frac{i \hbar t}{2mR^2}
\frac{\partial^2}{\partial \theta_i^2}\right]\otimes {\cal{I}}_2,
\end{equation}
where ${\cal{I}}_2$ is the two-dimensional identity matrix.
The trapping potential along the angular direction is
$V_{\mathrm{trap}}\left(\theta, t\right)=\frac{1}{2}m\omega_{\theta}^2 R^2\theta^2
\Theta(-t)$, and we assume that for the both components, the initial mean-field
wave function at $t=0$ is a Gaussian wave packet, i.e., ground state of the
harmonic trap,
$\Psi\left(\theta, 0\right)=\left(\frac{1}{\sqrt{\pi}\sigma}\right)^{\frac{1}{2}}
\mathrm{exp}\left\{-\frac{\left[\theta-\theta(0)\right]^2}{2\sigma^2}\right\}$ for
$\theta \in \left[\theta(0)-\pi, \theta(0)+\pi\right]$, where $\theta(0)=0$ and
$\sigma=\sqrt{\hbar/\left(m\omega_{\theta}\right)}/R \ll \pi$ are the initial
center and the width of the wave packet, respectively. Due to the periodicity of
the $\theta$ coordinate, the wave function outside this interval
can be defined via $\Psi\left(\theta+2J\pi, 0\right)=\Psi\left(\theta, 0\right)$,
with $J$ being an integer. The multiqubit initial GHZ state, is given by
\begin{equation}
\label{eq:InitialState}
\left|\tilde{\psi}\left(\theta_1, \theta_2, ..., \theta_N; 0 \right) \right\rangle
= \frac{1}{\sqrt{2}}\prod_{i=1}^N \Psi\left(\theta_i, 0\right)
\left(\left|{\bf 0} \right\rangle + \left|{\bf 1}\right\rangle \right),
\end{equation}
for which the normalization condition is
$1 = \int \mathrm{d}\theta_1 \mathrm{d}\theta_2 \cdot \cdot \cdot \mathrm{d}\theta_N
\left \langle\tilde{\psi}\left(\theta_1, \theta_2, ..., \theta_N; 0 \right) | \tilde{\psi}
\left(\theta_1, \theta_2, ..., \theta_N; 0\right) \right\rangle$.
The interferometer is launched at $t=0$ via kicking
the two components with $\pm v$ group velocity, respectively, as in Refs.~\cite{KandesarXiv2013,HainePRL2016}.
The kicking operator reads $\hat{K}(v)=\mathrm{exp}\left(\frac{i}{\hbar}L_k\sum_{j=1}^N \theta_j \sigma_{jz}\right)$,
which plays the role of a beam splitter, with $L_k=mRv$ being the kicking
angular momentum and $\sigma_{jz}$ being the Pauli $Z$ matrix of the $j$th
particle. Finally, at time $t=\tau$ when the two components are recombined
for the first time~\cite{TimeNote}, the full quantum state is given by
\begin{eqnarray}
\label{eq:ReadoutState}
\left|\tilde{\psi}\left(\theta_1, \theta_2, ..., \theta_N; \tau \right) \right\rangle
= \hat{K}^{\dagger}(v) \bigotimes_{i=1}^N \hat{U}_i\left(\tau\right) \hat{K}(v) \nonumber \\
\left|\tilde{\psi}\left(\theta_1, \theta_2, ..., \theta_N; 0 \right) \right\rangle,
\end{eqnarray}
where $\hat{U}_i\left(\tau\right)$ is the time evolution operator for
the $i$th qubit at $t=\tau$, which is given by Eq.~(\ref{eq:Uoperator}).
Note that for well-defining a Sagnac phase, we have firstly assume the
interaction ${\cal U}=0$ during the interrogation. Consequently, the quantum
state in the spin subspace after tracing out the orbital degrees of freedom
related to $\Psi_{\xi}\left(\theta, \tau\right)$ is given by
$|\psi(\tau)\rangle=\left(\mathrm{e}^{i\phi_S}|{\bf 0}\rangle+|{\bf 1}\rangle\right)/\sqrt{2}$
(up to a global phase factor), where (see Appendix~\ref{apped:Sagnacphase} for
detailed derivations)
\begin{eqnarray}
\label{eq:SagnacPhase}
\phi_S = \beta N \Omega \tau^2
\end{eqnarray}
is the multiparticle Sagnac phase, with $\beta=2mv^2/\left(\pi \hbar\right)$.
This expression for $\phi_S$ is equivalent to $N$ times the well-known single-particle
Sagnac phase $2m\Omega A/\hbar$, where $A=\pi R^2$ is the
area of the Sagnac interferometer, and for constant $v$ we have
$A= v^2 \tau^2/\pi$.

As a result, a Sagnac pure phase gate as an unitary operation mapping the
initial state of the qubits to the readout state is constructed, which in the
GHZ subspace spanned by $\left\{|{\bf 0}\rangle, |{\bf 1}\rangle\right\}$ reads
\begin{eqnarray}
\label{eq:PhaseGate}
U(\phi_S) = \mathrm{diag}\left[\mathrm{e}^{i\phi_S}, 1\right],
\end{eqnarray}
and the rotation frequency can be extracted from the interference signal of
the final state. Following standard quantum
metrological protocols, the above Sagnac phase encoding and rotation frequency
readout processes are repeated for $\nu=T/\tau$ times to reach a high precision,
where $T$ is the total resource time (See Fig.~\ref{fig:SI}). And the standard
deviation $\delta \hat{\Omega}$ for any unbiased estimator $\hat{\Omega}$ of
the rotation frequency is bounded from below by the quantum Cram\'er-Rao bound
(QCRB)~\cite{Helstrom,Holevo}, $\delta \hat{\Omega} \ge 1/\sqrt{\nu F}$,
where $F$ is the quantum Fisher information (QFI) with respect to $\Omega$ for
the readout state, which is an effective theoretical tool for assessing the
performance of various interferometry schemes for quantum sensing~\cite{HainePRL2016}.
Equivalently, we have $\delta \hat{\Omega} \sqrt{T} \ge 1/\sqrt{F/\tau}$.
For more introduction to quantum sensing and QFI, see Appendix~\ref{apped:QFI}.
From Eq.~(\ref{eq:SagnacPhase}), we see that the scale factor $\cal{S}$ of the
interferometer is ${\cal{S}} \propto N \tau^2$, and the noiseless optimal QFI is~\cite{LuoPRA2017}
\begin{eqnarray}
\label{eq:NoiselessQFI}
F_0=\left(\partial_{\Omega} \phi_S\right)^2=\left(2mNA/\hbar\right)^2,
\end{eqnarray}
which achieves the Heisenberg scaling and increases monotonically with
the area of the interferometer.

\vspace*{-1.5ex}
\section{Optimal sensitivity under independent decoherence}
\vspace*{-1.5ex}
\label{sec:SensitivityDecoherence}
In previous derivation of the
multiqubit Sagnac phase we have neglected the interaction and local field
fluctuations. Now we consider the qubit dephasing arising from such effects, which can
not be neglected in a realistic system~\cite{InteractionNote}. And for
shortcomings of the mean-field analysis in MWI with interaction effects, see
e.g., Refs.~\cite{AghamalyanNJP2015,DasPRA2015}. The atom-atom collision and
local fluctuations may cause a random shift $\delta \omega(t)$ of the energy
difference for each qubit, which can be formulated as a Gaussian
random process with zero mean and the correlation function~\cite{ScullyBook,SzankowskiPRA2014}
$\left \langle \delta \omega(t) \delta \omega(t^{\prime}) \right \rangle=2\gamma \delta(t-t^{\prime})$,
where $\delta(t-t^{\prime})$ is the Dirac function. The ensemble average leads
to the exponential dephasing of the single qubit~\cite{ScullyBook,PuriPRA1992,TWChenPRA2003}.
Considering the dephasing of the $N$-qubit system, the master equation for the
state $\varrho(t)$ in the phase-covariant frame can be written as~\cite{DephasingNote}
\begin{eqnarray}
\label{eq:Master}
\frac{\mathrm{d}\varrho(t)}{\mathrm{d}t}=\frac{\gamma}{2}\sum_{i=1}^N \left[\sigma_{iz}\varrho(t)\sigma_{iz}-\varrho(t)\right],
\end{eqnarray}
where $\gamma > 0$ is the dephasing strength.
\begin{figure}
\centerline{\includegraphics[height=2.2in,width=3.5in, clip=true]{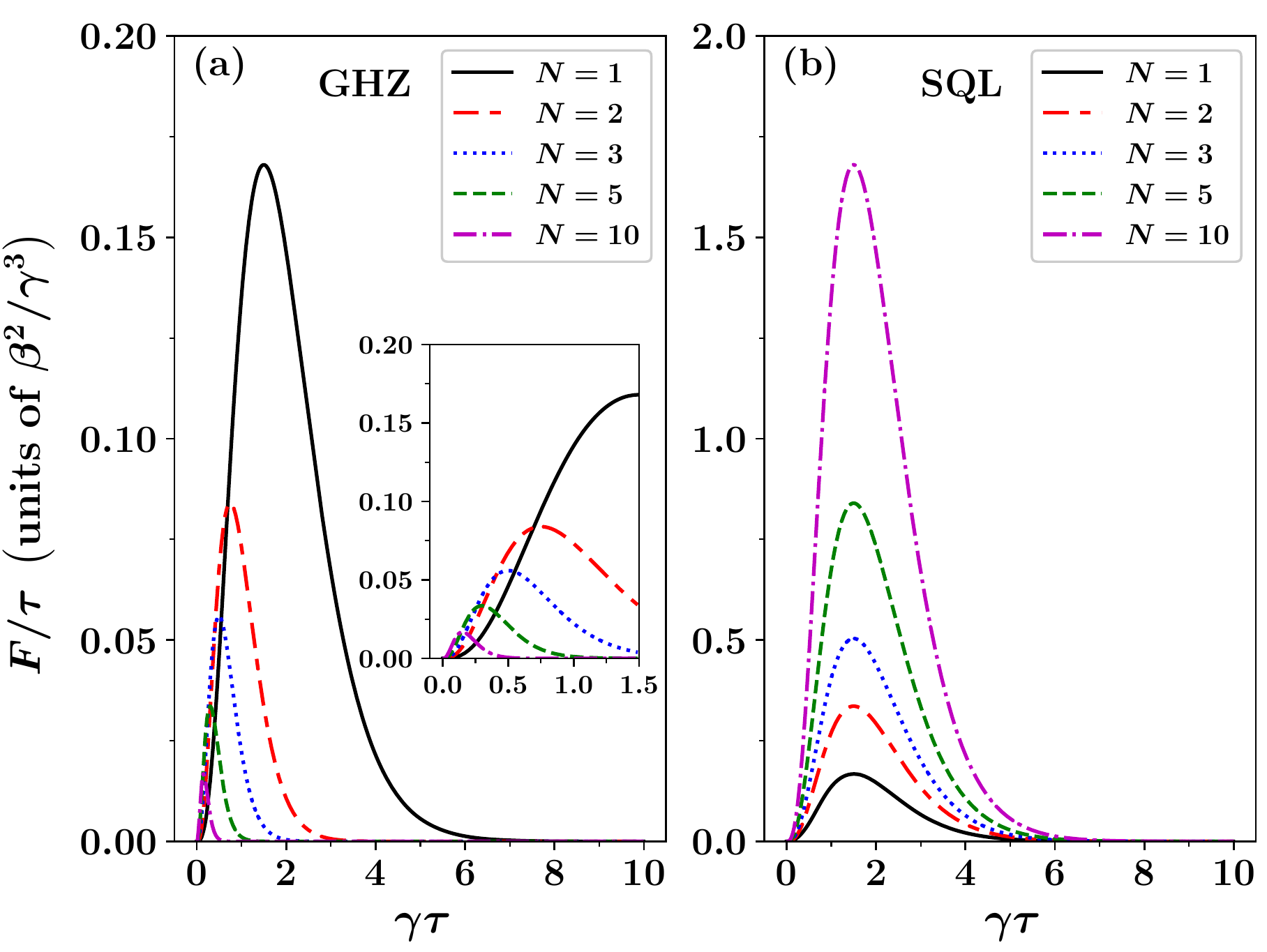}}
\caption{(Color online) The QFI $F/\tau$ (in units of $\beta^2/\gamma^3$) of Sagnac interferometers
as functions of $\gamma \tau$ for increasing the qubit number $N$ with (a) GHZ probe and (b)
uncorrelated qubits, where SQL represents the standard quantum limit and $F_{SQL}/\tau= N \beta^2 \tau^3 e^{-2 \gamma \tau}$.
The inset in (a) presents the detailed structure near the optimal interrogation time $\tau_{opt}$
and shares the same axes with the main panel (a). For panel (a), both the $\left(F /\tau\right)_{opt}$ and $\tau_{opt}$
are proportional to $N^{-1}$, and the QFI is completely lost in the large $N$ region. In panel (b),
the $\tau_{opt} = \mathrm{const}$ and $\left(F_{SQL} /\tau\right) \propto N$ (SQL). Note that the
black solid lines in panel (a) and (b) represent exactly the same function.}
\label{fig:QFI}
\end{figure}
\begin{table*}[t]
\centering
\caption{Comparison between Ramsey and Sagnac interferometers with GHZ probe states}

\begin{tabular}{c c c c c c c c c}
\hline \hline
Interferometers            & Quantity  &   $H_{\mathrm{single}}$    &   Phase   & Noiseless $F /\tau$ & Noiseless $\tau_{opt}$ & Dephasing & Noisy $\left(F /\tau\right)_{opt}$  & Noisy $\tau_{opt}$ \\ \hline
Ramsey                     & $\omega$  &  $\hbar \omega \sigma_z/2$   &$N \omega \tau$       &     ${\cal{O}}\left(N^2\right)$   & $T$ & $e^{-N\gamma \tau}$ &  ${\cal{O}}\left(N\right)$~\cite{HuelgaPRL1997,EscherNPhys2011,DemkowiczNC2012} & $1/\left(2N\gamma\right)$  \\ \hline
Sagnac                     & $\Omega$  &  $-\Omega \hat{L}_z$ & $\beta N \Omega \tau^2$   &${\cal{O}}\left(N^2\right)$~\cite{LuoPRA2017}    & $T$ & $e^{-N\gamma \tau}$ & ${\cal{O}}\left(N^{-1}\right)$ & $3/\left(2N\gamma\right)$ \\
\hline \hline
\end{tabular}
\label{tb:RamseySagnac}
\end{table*}
The completely positive and trace preserving (CPTP) map ${\cal E}$, which is a solution of
Eq.~(\ref{eq:Master}) and maps $\varrho_0$ to $\varrho(\tau)$,
is $\varrho(\tau)={\cal E}\left(\varrho_0\right)=\bigotimes_{i=1}^{N}{\cal E}_{i}\left(\varrho_0\right)$,
where $\varrho_0=\rho_0=|\psi_0\rangle\langle\psi_0|$ is the initial state and
${\cal E}_{i}(\varrho_0)=E_{i0}\varrho_0E_{i0}^{\dagger}+E_{i1}\varrho_0E_{i1}^{\dagger}$ is the local
phase-flip error operator for the $i$th qubit, with $E_{i0}=\sqrt{1-p(\tau)}{\cal{I}}_2$ and $E_{i1}=\sqrt{p(\tau)}\sigma_{iz}$
being the Kraus operators, where $p(\tau)=\left(1-e^{-\gamma \tau}\right)/2$
is the single-qubit error probability. Then it is straightforward to reach the readout state~\cite{ReadoutNotes},
$\rho(\tau) = {\cal E}\left[U(\phi_S)\rho_0 U^{\dagger}(\phi_S)\right] = \left[|{\bf 0}\rangle\langle{\bf 0}|+|{\bf 1}
\rangle\langle{\bf 1}|+(e^{-N \gamma \tau}e^{i\phi_S} |{\bf 0}\rangle\langle {\bf 1}|+\mathrm{h.c.})\right]/2$,
where $\phi_S$ is given by Eq.~(\ref{eq:SagnacPhase}). The QFI with respect to $\Omega$ for this
state is (see Appendix~\ref{apped:QFIReadOut})
\begin{eqnarray}
\label{eq:QFIwithNoise}
F = \beta^2 N^2 \tau^4 e^{-2N \gamma \tau}.
\end{eqnarray}
Note that here $F$ is the interrogation-time dependent QFI at the point where the Sagnac phase
accumulation is accomplished, which is actually $F_S$ in Ref.~\cite{HainePRL2016}. From
Eq.~(\ref{eq:QFIwithNoise}) we see that the optimal interrogation time and the Sagnac area $A$
are constrained by decoherence in noisy scenarios.

The precision bound for rotation sensing is determined by $F/\tau$ and its optimized value over
the interrogation time is given by
\begin{eqnarray}
\label{eq:QFI-opt}
\left(F/\tau\right)_{opt}=\beta^2\left(\frac{3}{2\gamma e}\right)^3 \frac{1}{N},
\end{eqnarray}
with $\tau_{opt}=3/(2N\gamma)$~\cite{MarkovianNote}. Therefore, the maximally entangled state reduces
the precision with increasing $N$ even under uncorrelated dephasing, which is a completely new result
in contrast to that was found in Ramsey-type interferometers~\cite{HuelgaPRL1997,EscherNPhys2011,DemkowiczNC2012}.
On the other hand, the QFI for uncorrelated qubits is given by $F_{SQL}= N \beta^2 \tau^4 e^{-2 \gamma \tau}$,
and is proportional to $N$ for any value of $\tau$. Shown in Fig.~\ref{fig:QFI} is the
interrogation-time normalized $F$ for increasing the qubit number $N$ with (a) GHZ probe and (b)
uncorrelated qubits, respectively. For a given value of $N$ and increasing $\tau$, the power law
$\tau^3$ behavior dominates at the beginning while the exponential prevails after reaching a maximum.
While the SQL is achieved in Fig.~\ref{fig:QFI}(b) with uncorrelated qubits, in contrast, the QFI
curves are shrinking with increasing $N$ in Fig.~\ref{fig:QFI}(a) for the GHZ probe. The physical
reason is that $\phi_S\left(\tau=\tau_{opt}\right) \propto N^{-1}$, such that the accumulated phase
signal is weakened with increasing $N$.

\vspace*{-1.5ex}
\section{Assessments of generic multiqubit MWI schemes}
\vspace*{-1.5ex}
\label{sec:GenericMWI}
Above results can be generalized
to more generic MWI schemes for quantum sensing with entangled states and under
independent dephasing, which were previously only considered in single-qubit or
noiseless scenarios. In general, for GHZ probes in the presence of independent
dephasing, if the accumulated phase is $\phi_{\chi} (\tau) \propto N \chi \tau^{\lambda}$
with $\chi$ being the physical quantity to be sensed and $\lambda > 0$ the time exponent
of the scale factor, then the optimal QFI with respect to $\chi$ is
\begin{eqnarray}
\left(F_{\chi}/\tau\right)_{opt}={\cal{O}}\left(N^{3-2\lambda}\right),
\end{eqnarray}
with $\tau_{opt}=(2\lambda-1)/(2N\gamma)$ (see Appendix~\ref{apped:QFIReadOut}).
Therefore, we may conclude that the Heisenberg scaling is actually inaccessible
because the condition $\tau_{opt} > 0$ requires $\lambda > 1/2$.
And for $\lambda \ge 1$, the best QFI can be achieved is $\left(F_{\chi}/\tau\right)_{opt}={\cal{O}}(N)$
(SQL) with the $\lambda=1$ class~\cite{HuelgaPRL1997,EscherNPhys2011,DemkowiczNC2012}.
For classes with $\lambda \ge 2$, the entangled probes could reduce the precision
with increasing the particle number.

Many of the current mainstream MWI schemes with atomic clock states belong to
the $\lambda=2$ class (without entanglement), and so do the proposed schemes
in Refs.~\cite{PangNC2017,GefenPRA2017} with time-dependently controlled
Hamiltonians. For example, the atom gravimetry considered in
Refs.~\cite{KasevichPRL1991,KasevichAPB1992,SchleichPRL2013,SchleichNJP2013,KritsotakisarXiv2017},
with the single-qubit phase $\phi_{{\bf g}}(\tau)={\bf{k}}_0 \cdot {\bf g} \left(\tau/2\right)^2$
and the atom free-propagation Sagnac interferometers in
Refs.~\cite{GustavsonPRL1997,GustavsonCQG2000,DurfeePRL2006,BarrettCRPhys2014},
with the encoded phase
$\phi_{{\bf \Omega}}(\tau)={\bf{k}}_0 \cdot \left({\bf{v}}_0 \times {\bf{\Omega}}\right) \tau^2/2$,
where $\bf g$ is gravitational acceleration, and ${\bf{v}}_0$ and ${\bf{k}}_0$
are the semiclassical velocity of atoms and effective Raman propagation
vectors, respectively. In Table~\ref{tb:RamseySagnac} we give a comparison
between standard Ramsey interferometers for atomic clocks ($\lambda=1$) and
Sagnac interferometers for rotation sensing ($\lambda=2$)~\cite{SagnacNote},
with GHZ input states and $H_{\mathrm{single}}$ denotes the single-particle
sensing Hamiltonian.

\vspace*{-1.5ex}
\section{QFI in the presence of collective dephasing}
\vspace*{-1.5ex}
\label{sec:CollectiveDephasing}
For closely spaced atoms in a Bose-Einstein condensate, they may collectively
couple to the external-field fluctuations~\cite{FerriniPRA2011,DornerNJP2012,ZhongPRA2013,SzankowskiPRA2014}.
Here we give the scaling behavior the QFI in the presence of collective
dephasing for matter-wave interferometers with GHZ probe and different scale
factors.

The master equation for the state $\varrho(t)$ in the phase-covariant frame in
the presence of collective dephasing is given by~\cite{DornerNJP2012,ZhongPRA2013}
\begin{eqnarray}
\label{eq:SMaster}
\frac{\mathrm{d}\varrho(t)}{\mathrm{d}t}=\Gamma \left[2J_z\varrho(t)J_z-J_z^2\varrho(t)-\varrho(t)J_z^2\right],
\end{eqnarray}
where $\Gamma > 0$ is the collective dephasing strength and $J_z=\sum_{i=1}^N \sigma_{iz}/2$
is the third component of the collective spin operator. For the GHZ state, the
readout state at time $t=\tau$ can be analytically calculated in a similar way,
which is given by
\begin{equation}
\rho(\tau)=\frac{1}{2}\left[|{\bf 0}\rangle\langle{\bf 0}|+|{\bf 1}
\rangle\langle{\bf 1}|+(e^{-N^2 \Gamma \tau}e^{i\phi(\tau)} |{\bf 0}\rangle\langle {\bf 1}|+\mathrm{h.c.})\right],
\end{equation}
where $\phi$ is the multiparticle interferometer phase
$\phi(\tau) \propto N \chi \tau^{\lambda}$ with the scale factor
${\cal{S}} \propto N \tau^{\lambda}$. The QFI with respect to $\chi$ for this
state can be calculated with the same method as in Appendix~\ref{apped:QFIReadOut}
and is given by
\begin{eqnarray}
{\cal{F}}_{\chi} &=& \left[\frac{\partial \phi (\tau)}{\partial \chi}\right]^2 e^{-2N^2 \Gamma \tau}
= {\cal{S}}^2 e^{-2N^2 \Gamma \tau} \nonumber \\
    &\propto& N^2 \tau^{2\lambda} e^{-2N^2 \Gamma \tau}.
\end{eqnarray}
And the the interrogation-time optimized value of ${\cal{F}}_{\chi}/\tau$ is
\begin{equation}
\left({\cal{F}}_{\chi}/\tau\right)_{opt} \propto N^{4\left(1-\lambda\right)} \left( \frac{2\lambda-1}{2\Gamma e}\right)^{2\lambda-1},
\end{equation}
with $\tau_{opt}=(2\lambda-1)/(2N^2 \Gamma)$. So for Ramsey-type interferometers
($\lambda =1$), the optimal QFI is
$\left({\cal{F}}_{\chi}/\tau\right)_{opt}=1/\left(2\Gamma e\right) = \mathrm{const}$~\cite{DornerNJP2012,ZhongPRA2013,SzankowskiPRA2014}
and is independent of $N$. While for Sagnac-type interferometers with $\lambda =2$,
the decoherence time is $\tau_{opt} \propto N^{-2}$ and
$\left({\cal{F}}_{\chi}/\tau\right)_{opt} \propto N^{-4} \left[3/\left(2\Gamma e\right)\right]^3$,
which decreases very rapidly with increasing the particle number $N$. This is
in stark contrast to the constant precision of Ramsey-type
interferometers~\cite{DornerNJP2012,ZhongPRA2013,SzankowskiPRA2014}.

\vspace*{-1.5ex}
\section{Recovering the Heisenberg scaling with quantum error correction}
\vspace*{-1.5ex}
\label{sec:QEC}
To tentatively recover the Heisenberg scaling and improve the
sensitivity, we theoretically explore in the following the potential of quantum
error-correction (QEC) codes for MWI schemes under local dephasing. QECs have been realized in
experiments for quantum computation~\cite{SchindlerScience2011,WaldherrNature2014,OfekNature2016,UndenPRL2016}
and have been proposed for quantum
metrology~\cite{DurPRL2014,KesslerPRL2014,ArradPRL2014,LuNC2015,OzeriArxiv2013,ReiterNC2017,DemkowiczPRX2017,ZhouNC2018}.
Here we analyze a QEC scheme with logical GHZ states proposed in
Refs.~\cite{DurPRL2014,LuNC2015}, which utilizes redundant qubits to suppress
phase-flip errors, with a possible application in the Sagnac atom interferometers.
As in Refs.~\cite{DurPRL2014,LuNC2015}, with $n$ physical qubits in each logical block,
the error probability $p(\tau)$ is exponentially suppressed by replacing the raw
GHZ state with a logical one~\cite{DurPRL2014,LuNC2015}, where the coding space
$\cal{C(G)}$ is stabilized by the stabilizer group $\cal{G}$. The $n$-qubit
($n$ is odd) phase-flip code is defined as ${\cal{C}}_n =\left\{|0\rangle_L, |1\rangle_L\right\}$,
where $|0\rangle_L =\left(|+\rangle^{\otimes n}+|-\rangle^{\otimes n}\right)/\sqrt{2}$
and $|1\rangle_L =\left(|+\rangle^{\otimes n}-|-\rangle^{\otimes n}\right)/\sqrt{2}$
are the bases for each logical qubit block, with $|+\rangle$ ($|-\rangle$) being
the eigenstate of the Pauli matrix $\sigma_x$ with eigenvalue $+1$ ($-1$). The
above code is stabilized by the operator $X_{\alpha}=\prod_{i \in \alpha} \sigma_{ix}$,
with $\alpha \subset \{1,2,3,...n\}$ and $|\alpha|=\mathrm{even}$, and is
capable of correcting $(n-1)/2$ phase-flip errors $\{\sigma_{iz}\}$~\cite{LuNC2015,DurPRL2014}.
With $N$ total physical qubits as resources, the number of logical qubits is
$N/n$. Furthermore, the error probability is renormalized to the logical level as
\begin{equation}
p_L(\tau)=\sum_{k=0}^{(n-1)/2} \dbinom{n}{k} p^{n-k}(\tau)\left[1-p(\tau)\right]^k.
\end{equation}
The QFI can be rewritten in terms of $p(\tau)$ using $e^{-\gamma \tau}=1-2p(\tau)$,
and the logical QFI in terms of $p_L(\tau)$ is given by
\begin{eqnarray}
F_L &=& \left(n\beta\right)^2 \left(N/n\right)^2 \tau^4 \left[1-2p_L(\tau)\right]^{2N/n} \nonumber \\
          &=& \beta^2 N^2 \tau^4 \left[1-2p_L(\tau)\right]^{2N/n}.
\end{eqnarray}
Now the quantum Cr\'{a}mer-Rao bound for the rotation frequency sensing is
$\delta \Omega \sqrt{T}=1/\sqrt{F_L/\tau}$.
\begin{figure}
\centerline{\includegraphics[height=2in,width=3.5in, clip=true]{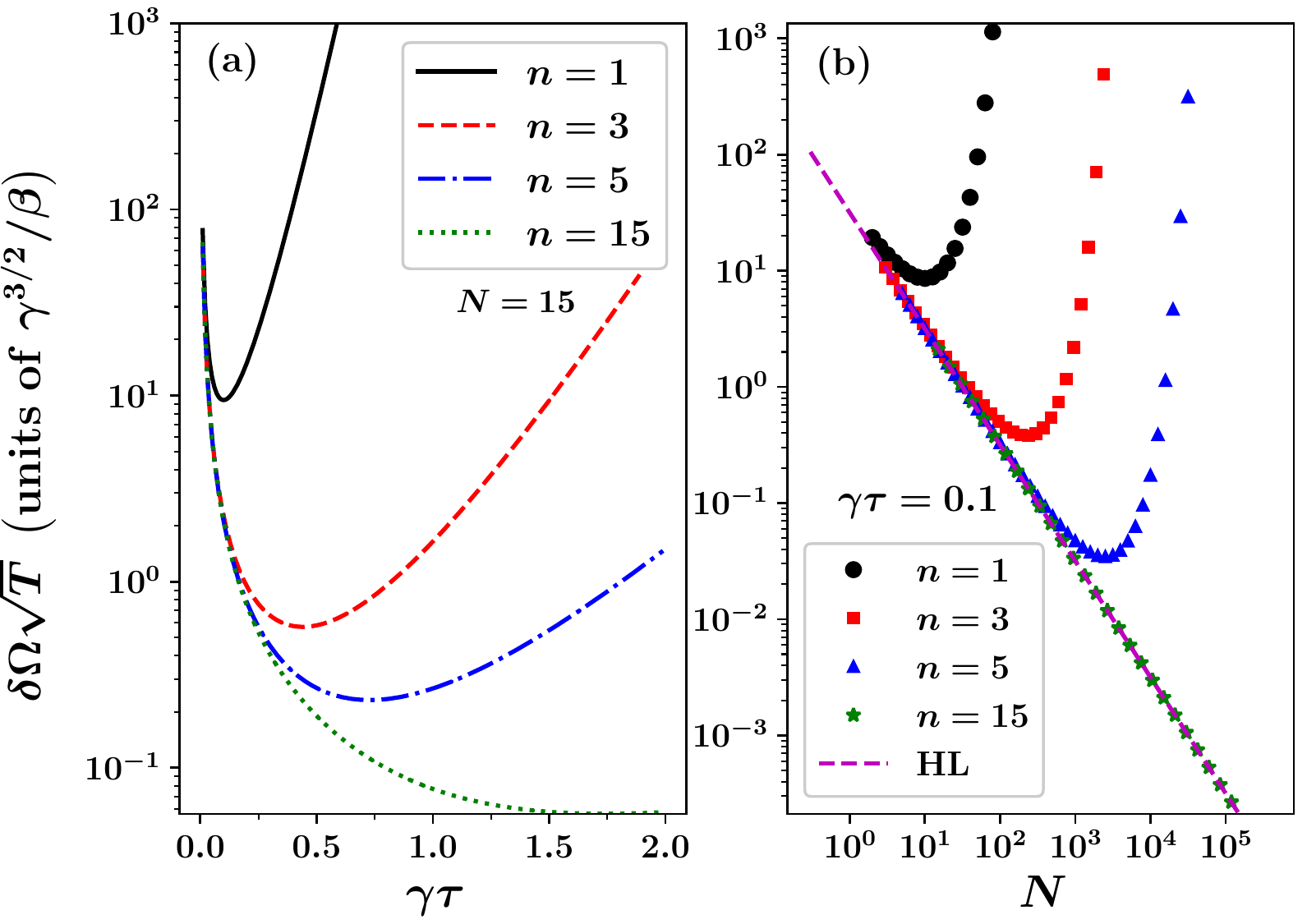}}
\caption{(Color online) The QCRB $\delta \Omega \sqrt{T}=1/\sqrt{F_L/\tau}$
(in units of $\gamma^{3/2}/\beta$) for rotation frequency sensing.
(a) $\delta \Omega \sqrt{T}$ vs $\gamma \tau$ for fixed total qubit number $N=15$ and
(b) $\delta \Omega \sqrt{T}$ vs $N$ for $\gamma \tau =0.1$,
with increasing the number of qubits $n$ in each logical block. HL in panel (b) denotes the
Heisenberg limit (magenta dashed). See the main text for details.}
\label{fig:QCRB}
\end{figure}

Plotted in Fig.~\ref{fig:QCRB}(a) is $\delta \Omega \sqrt{T}$ vs $\gamma \tau$ for a given
total qubit number $N=15$ and increasing qubit number $n$ in each logical block. With small
$p(\tau)$, the optimal interrogation time which minimizes the precision bound is $\tau_{opt}=3/\left(2N\gamma_{eff}\right)$,
as in Eq.~(\ref{eq:QFI-opt}), with the effective dephasing strength
$\gamma_{eff}=\gamma {\cal{O}}\left[p^{(n-1)/2}(\tau_{opt})\right]$. We find that the
use of the logical code will increase $\tau_{opt}$ in a power law fashion and improve
the sensitivity, where the preservation of Heisenberg scaling is promising.
Shown in Fig.~\ref{fig:QCRB}(b) is $\delta \Omega \sqrt{T}$ vs $N$ for $\gamma \tau=0.1$.
The representative values for $n$ are taken the same as that in the panel (a). One sees that for
each $n$, there exists an optimal total qubit number $N_{opt}$, where a minimum precision bound is
attained. Furthermore, for $N \in [1, N_{opt})$, the Heisenberg scaling (shown with magenta dashed)
is achieved~\cite{DurPRL2014}. For small $\gamma \tau$ and $p(\tau)$, it is straightforward to obtain
$N_{opt}=\mathrm{int}\left\{1/\left[2p_L + {\cal{O}} \left(p_L^2\right)\right]\right\} n$, where
$\mathrm{int}\{y\}$ denotes the integer part of $y$. For the set of values of $n$ and $\gamma \tau$
in Fig.~\ref{fig:QCRB}(b), $N_{opt} \approx (10, 219, 2320, 4.5 \times 10^7)$ for $n=(1, 3, 5, 15)$,
respectively. Therefore, with the help of the logical code, the effective scope for the Heisenberg
scaling can be extended.

\vspace*{-1.5ex}
\section{Conclusion and discussion}
\vspace*{-1.5ex}
\label{sec:Conclusion}
In summary, we have presented an assessment of the
optimal precision given by the QCRB for matter-wave interferometers, with
multiqubit GHZ input and in the presence of decoherence. Our
results show that due to the competition between the unconventional phase
accumulation (i.e., $\lambda \ge 2$) and the exponential dephasing, the use
of entangled probes leads to vanishingly small QFI while increasing the particle
number, which challenges the conventional wisdom. Finally, for completeness,
we tentatively analyzed a QEC scheme with logical GHZ states, which could have
the potential to protect the Heisenberg scaling.

It is worth noting that the non-entangled spin state and maximally entangled
GHZ state with unconventional interferometric scale factors are investigated
and compared in our work, due to the analytical computability of the QCRB for
such states. We show that the latter gives a much worse precision than the
former for Sagnac-type interferometers in the presence of uncorrelated
dephasing. Intuitively, there should be a maximal precision arising from the
balance between the entanglement-enhancement and noise-reduction effects. With
the general method in Ref.~\cite{EscherNPhys2011} for estimating the upper
bound of the noisy QFI maximized over all possible input states and by replacing
the scale factor ${\cal{S}} \propto \tau$ of Ramsey-type interferometers with
${\cal{S}} \propto \tau^2$, one can obtain that the use of (partially) entangled
states can only give $2.8\%$ relative precision enhancement with respect
to the uncorrelated spin state for Sagnac interferometers. This is quite minor
compared to the $\sqrt{e}$ enhancement in Ramsey-type
interferometers~\cite{HuelgaPRL1997,EscherNPhys2011,DemkowiczNC2012}.

\vspace*{-1.ex}
\acknowledgments
\vspace*{-1.5ex}
We acknowledge helpful discussions with Jun Xin and
Hui-Ke Jin. This work was supported by the National Key Research and Development
Program of China (No.~2017YFA0304202 and No.~2017YFA0205700), the NSFC through
Grant No.~11875231, and the Fundamental Research Funds for the Central Universities
through Grant No.~2018FZA3005. J. Liu acknowledges the support from the National
Natural Science Foundation of China (Grant No.11805073). X.M. Lu acknowledges
support from the National Natural Science Foundation of China under Grant
Nos.~61871162 and 11805048, and the Natural Science Foundation of Zhejiang Province
under Grant No.~LY18A050003.

%===============================End of Main Text====================================================
\appendix

\section{Derivation of the multiqubit Sagnac phase in Eq. (\ref{eq:SagnacPhase})}
\label{apped:Sagnacphase}
Here we provide detailed derivation of the multiqubit Sagnac phase in
Eq. (\ref{eq:SagnacPhase}) in the main text. After applying the kicking operator
$\hat{K}(v)$, the mean-field wave function for the $j$th
particle of $\left| \xi\right\rangle_j$ spin state ($\xi=0, 1$) is given by
$\Psi_{\xi}\left(\theta_j, 0\right)=\Psi\left(\theta_j, 0\right) \mathrm{exp}
\left[(-1)^\xi i L_k\theta_j/\hbar\right]$, which can be directly
obtained with $\sigma_{jz}\left| \xi\right\rangle_j = (-1)^{\xi}\left| \xi\right\rangle_j$,
and $\Psi\left(\theta_j, 0\right)$ is the initial Gaussian wave packet.
Therefore, the wave function at time $t$ reads
\begin{eqnarray}
\Psi_{\xi}\left(\theta_j, t\right) \otimes \left| \xi\right\rangle_j
= \hat{U}_j(t)\Psi_{\xi}\left(\theta_j, 0\right)
\otimes \left| \xi\right\rangle_j.
\end{eqnarray}
In addition, the Fourier transform of the initial Gaussian wave packet is given by
$\Psi\left(\theta, 0\right)=\left[1/(2\pi)\right]^{1/2}\sum_{l=-\infty}^{l=+\infty}
\tilde{\Psi}(l)\mathrm{exp}\left(il\theta\right)$, where
\begin{eqnarray}
\tilde{\Psi}(l)&=&\left[1/(2\pi)\right]^{1/2} \int_{-\pi}^{\pi} \Psi\left(\theta, 0\right)
\mathrm{exp}\left(-il\theta\right) \mathrm{d}\theta \nonumber \\
&=&\left(\sigma/\sqrt{\pi}\right)^{1/2} \mathrm{exp}\left(-\sigma^2 l^2/2\right)
\mathrm{erf}\left(\frac{\pi+i\sigma^2l}{\sqrt{2}\sigma}\right)  \nonumber \\
&\approx& \left(\sigma/\sqrt{\pi}\right)^{1/2} \mathrm{exp}\left(-\sigma^2 l^2/2\right),
\end{eqnarray}
where $\mathrm{erf}(z)=\frac{2}{\sqrt{\pi}}\int_{0}^{z}\mathrm{exp}\left(-t^2\right)\mathrm{d}t$
is the Gaussian error function, for which $\mathrm{erf}(z) \rightarrow 1$ when $\mathrm{Re}z \rightarrow +\infty$,
which is the situation with $\sigma \ll \pi$ here.
And by applying the time evolution operator $\hat{U}_j(t)$, one can obtain
\begin{eqnarray}
\label{eq:SMeanFieldWF}
\Psi_{\xi}\left(\theta_j, t\right)
&\approx& \left(\frac{1}{\sqrt{\pi}\tilde{\sigma}(t)}\right)^{\frac{1}{2}}
\mathrm{exp}\left\{-\frac{\left[\theta_j-\theta^{\left(\xi\right)}(t)\right]^2}
{2\sigma \tilde{\sigma}(t)}\right\} \nonumber \\
&\times&  \mathrm{exp}\left[(-1)^{\xi} \frac{i}{\hbar}L_k \left(\Omega t+\theta\right) \right]
\mathrm{exp}\left[\frac{-it L_k^2}{2\hbar mR^2}\right] \nonumber \\
&\times& \sum_{n=-\infty}^{+\infty}\mathrm{exp}\left\{2\pi i n \kappa-2\pi^2n^2
/\left[\sigma \tilde{\sigma}(t)\right]\right\},
\end{eqnarray}
where $\tilde{\sigma}(t)=\sigma+i\hbar t/\left(mR^2\sigma\right)$ and
$\theta^{\left(\xi\right)}(t)=\left[(-1)^{\xi}v/R-\Omega\right]t$, and
$\kappa=-i\left[\theta_j-\theta^{\left(\xi\right)}(t)\right]
/\left[\sigma \tilde{\sigma}(t)\right]$. Furthermore, under the condition
$|\tilde{\sigma}(t)| \ll \pi$ for $t \in [0, \tau]$, we have $\sum_{n=-\infty}^{+\infty}
\mathrm{exp}\left\{2\pi i n \kappa-2\pi^2n^2/\left[\sigma \tilde{\sigma}(t)\right]\right\}
=1+\sum_{n=-\infty, n \ne 0}^{+\infty}\mathrm{exp}\left\{2\pi i n \kappa-2\pi^2n^2
/\left[\sigma \tilde{\sigma}(t)\right]\right\} \approx 1$, and then we obtain
\begin{eqnarray}
\label{eq:SmodulusWF}
\left|\Psi_{\xi}\left(\theta_j, t\right)\right|^2
\approx \frac{1}{\sqrt{\pi}\left|\tilde{\sigma}(t)\right|}
\mathrm{exp}\left\{-\frac{\left[\theta_j-\theta^{\left(\xi\right)}(t)\right]^2}
{\left|\tilde{\sigma}(t)\right|^2}\right\}. \nonumber \\
\end{eqnarray}
Therefore, at time $t$ and under the condition $|\tilde{\sigma}(t)| \ll \pi$,
the wave function in Eq.~(\ref{eq:SMeanFieldWF})
describes Gaussian wave packets centered at $\theta^{\left(\xi\right)}(t)$,
i.e., propagating in group linear velocity $(-1)^{\xi}v-\Omega R$, for $\xi=0$
and $1$, respectively, and with the same width $\left|\tilde{\sigma}(t)\right|$.
The interrogation time (or collision time) $\tau$, at which the two centers of
the counter-propagating Gaussian wave packets are completely overlapped, is
given by
$\theta^{\left(0\right)}(\tau)-\theta^{\left(1\right)}(\tau)=2\pi$, or
equivalently, $\tau=\pi R/v$.

With above results, one can obtain the multiparticle readout state
$\left|\tilde{\psi}\left(\theta_1, \theta_2, ..., \theta_N; \tau \right) \right\rangle$
in Eq.~(\ref{eq:ReadoutState}) and the corresponding density matrix reads
$\tilde{\rho}\left(\theta_1, \theta_2, ..., \theta_N; \tau \right)
=\left|\tilde{\psi}\left(\theta_1, \theta_2, ..., \theta_N; \tau \right) \right\rangle
\left \langle\tilde{\psi}\left(\theta_1, \theta_2, ..., \theta_N; \tau \right) \right|$.
The reduced density matrix in the spin subspace after tracing out the
orbital degrees of freedom related to $\Psi_{\xi}\left(\theta, \tau\right)$
is given by
\begin{eqnarray}
\label{eq:SReducedDensityMatrix}
\rho(\tau) &=&  \int \mathrm{d}\theta_1 \mathrm{d}\theta_2 \cdot \cdot \cdot
\mathrm{d}\theta_N \tilde{\rho}\left(\theta_1, \theta_2, ..., \theta_N;
\tau \right) \nonumber \\
           &=& \frac{1}{2}\left[|{\bf 0}\rangle\langle{\bf 0}|+|{\bf 1}
           \rangle\langle{\bf 1}|+(e^{i\phi_S} |{\bf 0}\rangle\langle {\bf 1}|
           +\mathrm{h.c.})\right],
\end{eqnarray}
where
\begin{eqnarray}
\label{eq:SSagnacPhase}
\phi_S = \beta N \Omega \tau^2
\end{eqnarray}
is the multiparticle Sagnac phase, with $\beta=2mv^2/\left(\pi \hbar\right)$.
This expression for $\phi_S$ is equivalent to $N$ times the well-known
single-particle Sagnac phase $2m\Omega A/\hbar$, where
$A=\pi R^2$ is the area of the Sagnac interferometer, and for constant $v$
we have $A= v^2 \tau^2/\pi$. The corresponding spin-subspace quantum
state can be written as $|\psi(\tau)\rangle=\left(\mathrm{e}^{i\phi_S}|{\bf 0}\rangle
+|{\bf 1}\rangle\right)/\sqrt{2}$ (up to a global phase factor), with which
$\rho(\tau)$ can be given by $\rho(\tau)=|\psi(\tau)\rangle \langle\psi(\tau)|$.

\section{Quantum sensing and quantum Fisher information}
\label{apped:QFI}
Here we present a brief introduction to quantum sensing and QFI. The QFI plays a crucial role in quantum
metrology and quantum sensing. Our basic quantum resources for a SAIG include $N$ cold probe (two-level)
atoms (qubits), total sensing time $T$, single-round interrogation time $\tau$, and the controlling and
measurement devices. In a standard metrological scheme, the initial sate of the probe is prepared
at $\rho_0$ and followed by a dynamical evolution $\rho_0 \xrightarrow{\phi_{\chi} (t)} \rho_{\chi}$ ($\rho_{\chi} := \rho_{\chi}(t)$),
which encodes the quantity $\chi$ to be sensed into the relative phase $\phi_{\chi}(t)$ of qubits, and can be
read out by quantum measurements after a single-round time $t=\tau$. Within the total time $T$, the
number of repetitive rounds of sensing and measurement is $\nu=T/\tau$. The standard deviation for any unbiased
estimator $\hat{\chi}$ is bounded from below by the quantum Cram\'er-Rao bound (QCRB)~\cite{Helstrom,Holevo},
\begin{equation}
\label{eq:SQCRB}
\delta \hat{\chi} \ge 1/\sqrt{\nu F},
\end{equation}
where $F$ is the QFI at $t=\tau$, or equivalently,
\begin{equation}
\label{eq:SC-R bound}
\delta \hat{\chi} \sqrt{T} \ge 1/\sqrt{F/\tau}.
\end{equation}
Thus, finding the optimal input state and quantum measurement to maximize the QFI is a central problem in
high precision quantum sensing. In general, the QFI of $\chi$ associated with $\rho_{\chi}$ is defined by
$F=\mathrm{Tr}(\rho_{\chi} L^2)$~\cite{Helstrom,Holevo}, where $\mathrm{Tr}$ is the trace operation and $L$ is the symmetric
logarithmic derivative (SLD) operator, which is given by
\begin{equation}
\partial_{\chi} \rho_{\chi}=\left(\rho_{\chi} L + L \rho_{\chi}\right) /2.
\end{equation}

Usually, a signal accumulation process is a unitary quantum channel, which gives $\rho_{\chi}=U_{\chi}\rho_0U^{\dagger}_{\chi}$,
where $U_{\chi}$ is a time and $\chi$ dependent unitary operator. It has been shown that for a pure state in unitary
quantum channels, the QFI can be obtained from the variance of a Hermitian operator
${\cal{H}}=i\left(\partial_{\chi}U^{\dagger}_{\chi}\right)U_{\chi}$ in $\rho_0$,
with~\cite{BoixoPRL2007,TaddeiPRL2013,PangPRA2014,LiuSR2015}
\begin{equation}
\label{eq:F-H}
F=4(\langle{\cal{H}}^2\rangle-\langle{\cal{H}}\rangle^2),
\end{equation}
where $\langle O\rangle:=\mathrm{Tr}(\rho_0 O)$ for any operator $O$. For an ensemble of $N$ qubits as the
input state in a standard Ramsey experiment, the maximal QFI ($\propto N^2$) is obtained when
$\rho_0$ is the GHZ state~\cite{PangPRA2014}, and when the inputs are uncorrelated qubits, $F \propto N$.
So the GHZ state gives the Heisenberg scaling for the sensing precision while the uncorrelated inputs
leads to the SQL, according to Eq.~(\ref{eq:SC-R bound}). However, in the presence of noises, the unitary
quantum channel will be modified by errors, and the corresponding QFI will be reduced or even be lost.
As a result, the expected sensing precision may not be achieved.

A special case is taking the quantity $\chi$ to be the relative phase $\phi$ of the two interferometric modes,
and the unitary phase imprinting operator is given by $U_{\phi}=\mathrm{exp}\left(-i\phi J_z\right)$, where
$J_z=\sum_{i=1}^N \sigma_{iz}/2$ is half of the relative number operator between the two modes. And the
corresponding $\cal{H}$ in Eq.~(\ref{eq:F-H}) is ${\cal{H}}=J_z$. Therefore, the QFI in Eq.~(\ref{eq:F-H})
is exactly the variance of the relative number with respect to the initial probe state, and the QCRB in
Eq.~(\ref{eq:SQCRB}) manifests itself as the uncertainty relation between the \emph{relative} phase and the
\emph{relative} number (take $\nu =1$). So the initial state with the largest relative number fluctuation
(e.g., the GHZ state) gives the highest phase resolution, while the total number of the state can be fixed.

\section{Calculations of the noisy QFI under independent dephasing}
\label{apped:QFIReadOut}
Here we give the detailed Calculations of the noisy QFI under independent dephasing and generalize the result
for Sagnac-type interferometers to more genetic classes. The spectral decomposition of the density matrix
$\rho$ is given by
\begin{equation}
\rho = \sum_{i=1}^d p_i \left|\psi_i \right\rangle \left\langle \psi_i \right|,
\end{equation}
where $d$ is the dimension of the support set of $\rho$, and $p_i$ is the $i$th eigenvalue of $\rho$,
with $\left|\psi_i \right\rangle$ being the corresponding $i$th eigenvector. With this
representation, the QFI with respect to the quantity $\chi$ can be expressed as~\cite{ZhangPRA2013,LiuCTP2014}
\begin{eqnarray}
\label{eq:SQFIExpression}
F = \sum_{i=1}^d \frac{\left(\partial_{\chi} p_i\right)^2}{p_i}
     + \sum_{i=1}^d 4p_i \left\langle \partial_{\chi} \psi_i | \partial_{\chi} \psi_i \right\rangle  \nonumber \\
    - \sum^d _{i,j=1; \atop p_i+p_j \ne 0} \frac{8p_i p_j}{p_i + p_j}\left|\left\langle \psi_i | \partial_{\chi} \psi_j \right\rangle \right|^2.
\end{eqnarray}
For the readout GHZ state
\begin{equation}
\rho(\tau)=\left[|{\bf 0}\rangle\langle{\bf 0}|+|{\bf 1}
\rangle\langle{\bf 1}|+(e^{-N \gamma \tau}e^{i\phi_S} |{\bf 0}\rangle\langle {\bf 1}|+\mathrm{h.c.})\right]/2
\end{equation}
at $t=\tau$, where $\phi_S = \beta N \Omega \tau^2$ is the Sagnac phase,
the dimension of the support set is $d=2$, with the two eigenvalues
$p_{\pm}=\left(1\pm e^{-N\gamma\tau}\right)/2$ and the corresponding eigenvectors
$\left|\psi_{\pm} \right\rangle =\left(e^{i\phi_S}|{\bf 0}\rangle \pm |{\bf 1}\rangle\right)/\sqrt{2}$,
respectively. The QFI with respect to the rotation frequency $\Omega$ can be readily calculated
from Eq.~(\ref{eq:SQFIExpression}) and is given by
\begin{eqnarray}
F &=& \left(\frac{\partial \phi_S}{\partial \Omega}\right)^2 e^{-2N \gamma \tau} \nonumber \\
    &=& \beta^2 N^2 \tau^4 e^{-2N \gamma \tau},
\end{eqnarray}
\begin{figure}[t]
\centerline{\includegraphics[height=2.5in,width=3.2in, clip=true]{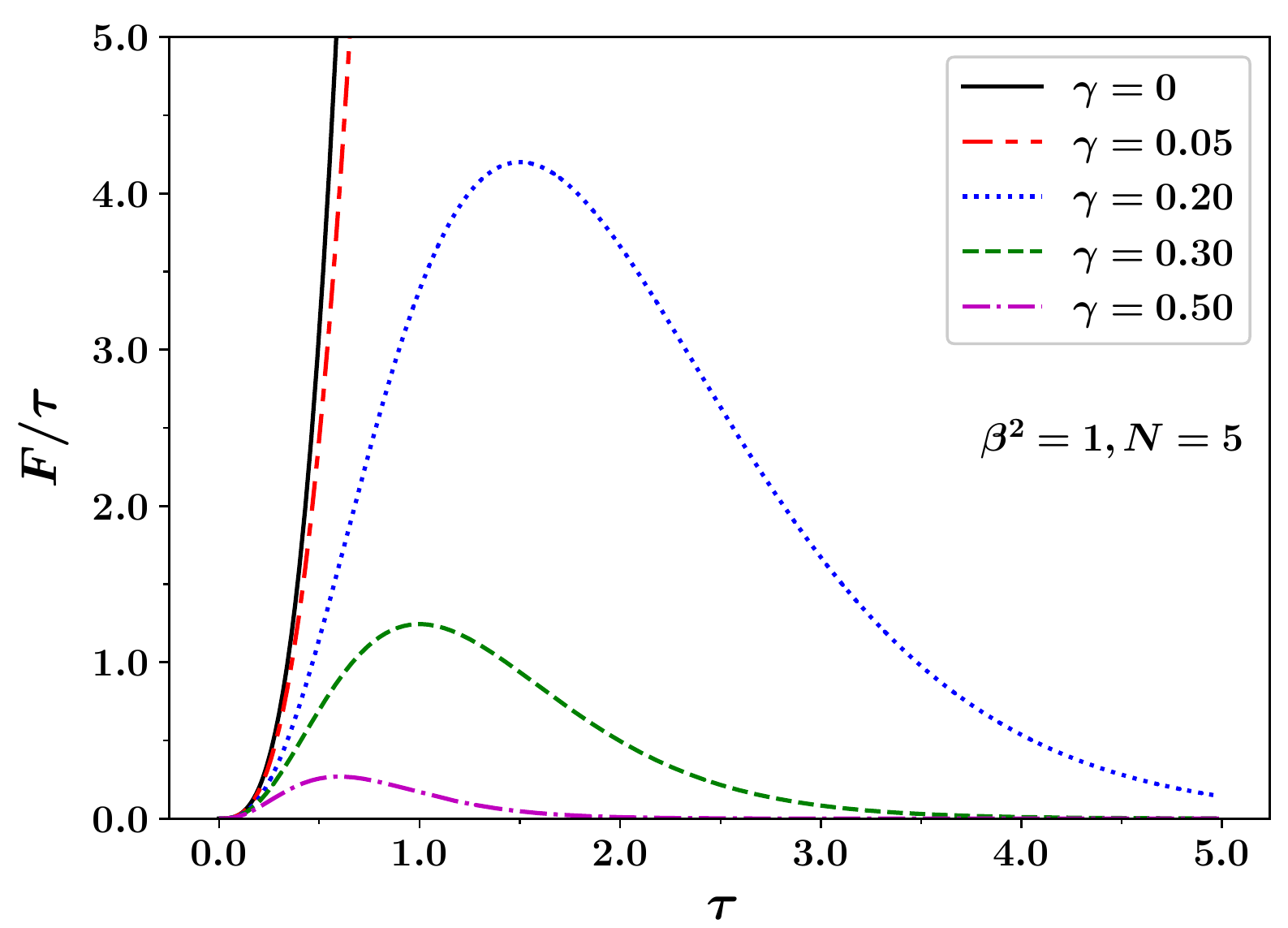}}
\caption{(Color online) Effects of increasing the dephasing strength $\gamma$ on $F/\tau$ vs $\tau$.
We set $\beta^2 = 1$ and the representative value for the qubit number is $N=5$.}
\label{Sfig:QFI}
\end{figure}
which is Eq.~(\ref{eq:QFIwithNoise}) in the main text. In noiseless scenarios
with $\gamma=0$, $F/\tau$ increases monotonically with $\tau$ while it has an
optimum for finite dephasing strength. In Fig.~\ref{Sfig:QFI} we show the effects
of increasing the dephasing strength on the interrogation-time normalized QFI of
the readout state with a fixed qubit number $N$. One sees that both of the optimal
$F/\tau$ and optimal interrogation time are decreasing with increasing $\gamma$.

\emph{QFI of generic MWI schemes.} For generic MWI schemes with GHZ states and under independent
dephasing noises, if the accumulated phase is $\phi_{\chi}(\tau) \propto N \chi \tau^{\lambda}$
($\lambda > 0$), then following the same procedure one can easily obtain the QFI with respect to the
quantity $\chi$, which is given by
\begin{eqnarray}
F_{\chi} &=& \left(\frac{\partial \phi_{\chi}}{\partial \chi}\right)^2 e^{-2N \gamma \tau} \nonumber \\
    &\propto& N^2 \tau^{2\lambda} e^{-2N \gamma \tau}.
\end{eqnarray}
And the the interrogation-time optimized value of $F_{\chi}/\tau$ is
$\left(F_{\chi}/\tau\right)_{opt}={\cal{O}}(N^{3-2\lambda})$, with $\tau_{opt}=(2\lambda-1)/(2N\gamma)$.
So for $\lambda \ge 1$, the best QFI can be achieved is $\left(F_{\chi}/\tau\right)_{opt}={\cal{O}}(N)$
(SQL) with the $\lambda=1$ class. See the main text.

%=============================== Reference=================================================
%\bibliography{Ref}

\begin{thebibliography}{75}
\bibitem [{Tim()}]{TimeNote}%
  \BibitemOpen
  \href@noop {} {}\bibinfo {note} {The interrogation time $\tau$ is defined at
  which the center of the two counter-propagating Gaussian wave packets are
  completely overlapped (See Appendix~\ref{apped:Sagnacphase}).}\BibitemShut {Stop}%
\bibitem [{Rea()}]{ReadoutNotes}%
  \BibitemOpen
  \href@noop {} {}\bibinfo {note} {Applying the $\pi/2$ pulse does not change
  the QFI with respect to $\Omega$, so here we refer the state just after the
  Sagnac phase encoding as the readout state.}\BibitemShut {Stop}%
\bibitem [{Sag()}]{SagnacNote}%
  \BibitemOpen
  \href@noop {} {}\bibinfo {note} {Note that if one fixes the initial kicking
  angular momentum $L_k$ while doing the optimization, then the scheme will
  belongs to the $\lambda=1$ class, as in
  Ref.~\cite{HalkyardPRA2010}.}\BibitemShut {Stop}%
\end{thebibliography}
%

\end{document}